# A Histopathology Study Comparing Contrastive Semi-Supervised and Fully Supervised Learning

Lantian Zhang[1,2], Mohamed Amgad[2], Lee A.D. Cooper[2]
[1] North Shore Country Day, Winnetka, IL, USA
[2] Department of Pathology, Northwestern University, Chicago, IL, USA

**Abstract**

*Data labeling is often the most challenging task when developing computational pathology models. Pathologist participation is necessary to generate accurate labels, and the limitations on pathologist time and demand for large, labeled datasets has led to research in areas including weakly supervised learning using patient-level labels, machine assisted annotation and active learning. In this paper we explore self-supervised learning to reduce labeling burdens in computational pathology. We explore this in the context of classification of breast cancer tissue using the Barlow Twins approach, and we compare self-supervision with alternatives like pre-trained networks in low-data scenarios. For the task explored in this paper, we find that ImageNet pre-trained networks largely outperform the self-supervised representations obtained using Barlow Twins.*

## 1 INTRODUCTION

### 1.2 Semi-supervised learning

Self-supervised learning (SSL) techniques have emerged as a popular approach for learning highly predictive representations from sparsely labeled datasets [1]. SSL approaches have yielded results that are competitive with supervised learning methods in vision applications using only a fraction of the labeled data used in fully-supervised approaches [2]. SSL representations can be trained with a variety of pretext tasks including context prediction or comparing learned feature representations of distorted versions of the same instance [3], [4], [5], [6]. SSL representations can then be combined with simpler fully supervised models or fine-tuned using small, labeled datasets for specific supervised tasks. The vast abundance of unlabeled data in pathology applications, as well as the scarcity of labels makes SSL particularly relevant for computational pathology problems [7]. SSL has been explored in pathology using pretext tasks, including predicting magnification or using color-deconvolved transformations for self-supervision [8]. Unfortunately, the best practices and opportunities for SSL in computational pathology are not fully understood. Fundamental SSL research publications use benchmarks like ImageNet classification that are more complex than narrow computational pathology tasks. Open questions include how well SSL findings using general benchmarks translate to computational pathology tasks, and how SSL performs compared to popular alternatives like pre-trained network representations.

In this paper we explore the Barlow Twins (BT) SSL method in the context of tissue classification in breast cancer [3]. We explore three questions: 1) How does BT performance compare to fully supervised networks with limited labeled data; 2) What hyperparameters and transformations are good for training BT representations for histopathology applications; 3) How does BT performance compare to conventional pre-trained models.

## 2 MODEL ARCHITECTURE

### 2.1 Barlow Twins

Barlow Twins (BT) is a SSL technique proposed by Zbontar *et al.* in 2021 [3]. BT uses a cross-correlation loss function to encourage similarity between the learned representations of two distorted "views" of an image. This loss encourages the encoder to produce a representation that is robust to distortion, and to reduce redundancy between features within that representation. Distortion transformations utilized in the original BT paper include Gaussian blurring, solarization, and color jittering. BT addresses problems with other SSL approaches that seek to learn distortion-invariant representations but that are sensitive to batch size or that often experience collapse and produce trivial representations.

Our implementation of the Barlow Twins model and its preprocessing pipeline follows the original paper and uses a ResNet-50 network as the encoder backbone [9]. This feeds a projector network with output dimension 4096. We chose not to increase the projector dimensionality beyond this due to computing power constraints. For classification we add a fully connected layer with softmax activation on top of the ResNet-50 backbone to output classification scores. We discard the projector network in the classification model, then fine-tune the pre-trained backbone together with the classification layer.

## 3 EXPERIMENTAL SETUPS

### 3.1 Materials

Our experiments use a publicly available dataset containing over 20,000 segmented tissue regions from 151 TCGA breast cancer slides at 0.25 MPP resolution [10]. Using these images and segmentation masks, we generated two datasets of tissue patches, one unlabeled and the other labeled. For the unlabeled dataset, we crop patches using a 224×224 pixel sliding window with a stride size of 112 pixels. This resulted in a total of 179,926 overlapping patches. Note that the segmentation masks were not utilized in this processing. We generate the labeled dataset in a similar manner. However, we further label each patch as one of six classes if over 30% of the pixels belong to a class according to the image's segmentation mask, for both train and test data. The classification labels were tumor, TILs, fat, stroma, necrotic debris, or plasma cell infiltrate. We discarded any patch that did not meet the 30% threshold or belonged to a class outside of the six selected. This resulted in a total of 139,053 labeled, overlapping patches. A label of adipose tissue was not included in the original dataset, and so we manually reviewed patches labeled as white space and separated the adipose patches from background.

The resulting datasets have a significant class imbalance, with the tumor and stroma classes comprising over three-quarters of the patches (**Figure 1**). Rare classes are a significant opportunity for methods like SSL that seek to do "less with more." Labeling rare classes can be very inefficient, requiring pathologists to search through large areas for rare phenomena. To highlight this imbalance, we report raw accuracy values for the four minor classes (plasma cell infiltrate, fat, necrotic debris, and TILs dense) during evaluation as "minor accuracy."

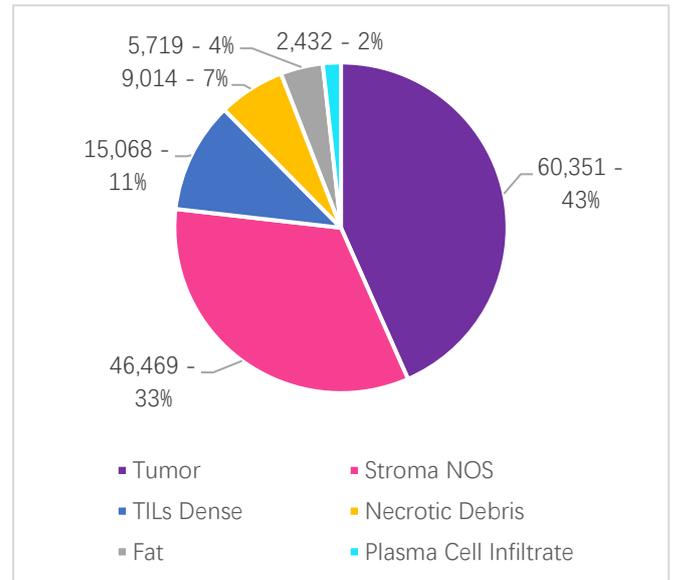

**Figure 1: Class distribution in labeled dataset.**

Finally, we separate training and testing instances at the level of the medical labs that produced the slides to evaluate generalizability. Patches in the labeled dataset from the following hospitals were used as the testing set: UCSF (A1), Walter Reed (A2), IUPUI (GI), and Ontario Institute for Cancer Research (HN). These hospitals provided ~15% of the slides included in the dataset. Within the rest of the labeled dataset (85% of total), a fixed 20% was used as a validation set while 80% was used to create training sets of varying sizes (**Table 1**). We stratified all splits to ensure that the training and validation sets had similar label distributions.

**Table 1: Training set sizes.**

| % Labeled Data | # Patches |
|---|---|
| 0.5 | 590 |
| 1 | 1182 |
| 2 | 2367 |
| 5 | 5920 |
| 10 | 11838 |
| 80 (max.) | 94715 |

### 3.2 BT training procedures

We used a two-stage training procedure to produce each BT classifier. In the pre-training stage, we trained the ResNet50 feature extractor and projection layer using the cross correlation loss for 100 epochs with a batch size of 256, learning rate of 1e-3, cosine decay schedule, and LAMB optimizer with weight decay of 1.5e-6 [11].

We then removed the projector from the pre-trained network and added a classification layer, retaining the pre-trained ResNet50 backbone as a feature extractor. Then, we fine-tuned the entire model, both the backbone and classification layer, for 30 epochs with a batch size of 256, uniform learning rate of 0.03, cosine decay schedule, and SGDW optimizer with categorial cross entropy [12]. Finally, we used stopping criteria to select the model with the highest micro-averaged AUROC score on the validation set.

Following Zbontar *et al.*'s observation that larger batch sizes lead to better performance, we used the largest batch size that would fit in our NVIDIA Tesla V100 GPUs [3]. The other hyperparameters were determined by a non-exhaustive manual search process with feedback from the validation set.

### 3.2 Fully supervised baseline models

We use two different baseline models for comparison. FS-scratch is a ResNet50 architecture with a classification layer trained from scratch using categorical cross entropy loss. This model did not utilize ImageNet weight initialization. FS-scratch models were trained for 30 epochs using a batch size of 256, learning rate of 5e-3, and Adam optimizer [13]. The learning rate was multiplied by a factor of 0.3 every time the micro-averaged validation AUROC plateaued for three consecutive epochs. We used the same stopping criteria as for the BT models. FS-ImageNet is the second baseline model; it uses ImageNet weight initialization and is otherwise identical to FS-scratch. We did not utilize augmentation techniques when training either of these baseline models.

## 4 RESULTS AND DISCUSSION

### 4.1 Barlow Twins and baseline comparison

**Table 2** and **Figure 2** present comparisons of the BT and fully supervised models with various amounts of labeled training data. Interestingly, we found that FS-ImageNet has superior AUROC performance in almost all cases. Performance of the FS-ImageNet models saturates quickly, starting at 0.937 micro-AUROC for 0.5% of labeled training instances, to 0.966 within 0.01 of peak micro-AUROC at 5% of labeled training instances (0.975). FS-ImageNet is consistently higher than FS-scratch in all cases except when 80% of labeled instances is used, where FS-scratch is slightly higher (0.980 versus 0.975).

Compared to FS-scratch, BT achieves superior AUROC performance when up to 5% of labels are used. BT had superior minor accuracy over both FS-ImageNet and FS-scratch on minor accuracy in this low-data regime as well.

The superior performance of FS-ImageNet is probably due in part to the relatively simplicity of the tissue classification task. In tasks where performance saturates with smaller amounts of data, an initialization strategy like ImageNet pre-training may be hard to outperform. Features learned in the general pattern recognition tasks are likely a superset of what is needed for good performance on specific histopathology tasks such as tissue classification. In fact, a varied dataset such as TCGA, that is sourced from multiple medical labs, may facilitate the development of robust trained-from-scratch classifiers like FS-ImageNet. The performance of FS-scratch also suggests that enough label data exists to saturate performance on this task with a small subset of labels.

Although further investigation is needed, it is possible that BT and other SSL methods may not provide the same benefits in computational pathology applications as they do in general-purpose SSL benchmarking applications. Given that the performance crossover may be relatively low in simple applications, BT may be most useful as a starting point for methods like assisted annotation, until enough labels are acquired to switch to a full supervision.

### 4.3 Impact of hyperparameters and distortion transformations

We experimented with a variety of BT hyperparameters for our training setup. First, we modified the projector network dimension of the BT model during the first stage of training. Unlike what Zbontar *et al.* suggests, we do not observe a significant positive correlation between projector dimensionality and performance [3]. This could be because we use fewer data during pre-training or the fact that we did not explore dimensions greater than 4096.

**Table 2: Performance of BT and supervised models**

| Method | % Labeled data | AUROC (micro) | AUROC (macro) | Minor Acc. |
|---|---|---|---|---|
| FS-scratch | 0.5 | 0.4858 | 0.5000 | 0.6491* |
| | 1 | 0.6217 | 0.5422 | 0.0450 |
| | 2 | 0.9178 | 0.8539 | 0.1256 |
| | 5 | 0.9469 | 0.8967 | 0.4625 |
| | 10 | 0.9542 | 0.9132 | 0.4701 |
| | 80 | **0.9800** | **0.9625** | **0.7626** |
| BT-pre-train | 0.5 | 0.9073 | 0.8532 | **0.6344** |
| | 1 | 0.9435 | 0.8916 | 0.6331 |
| | 2 | 0.9534 | 0.9099 | 0.5707 |
| | 5 | 0.9610 | 0.9243 | 0.6190 |
| | 10 | 0.9611 | 0.9241 | 0.6588 |
| | 80 | 0.9681 | 0.9460 | 0.6748 |
| FS-ImageNet | 0.5 | **0.9368** | **0.8933** | 0.5143 |
| | 1 | **0.9514** | **0.9284** | **0.6717** |
| | 2 | **0.9638** | **0.9392** | **0.7104** |
| | 5 | **0.9656** | **0.9586** | **0.7159** |
| | 10 | **0.9670** | **0.9594** | 0.6632 |
| | 80 | 0.9746 | 0.9510 | 0.7332 |

\* Resulted from constant predictions of "TILs" regardless of input data; we opt not to include this statistic in **Figure 2**

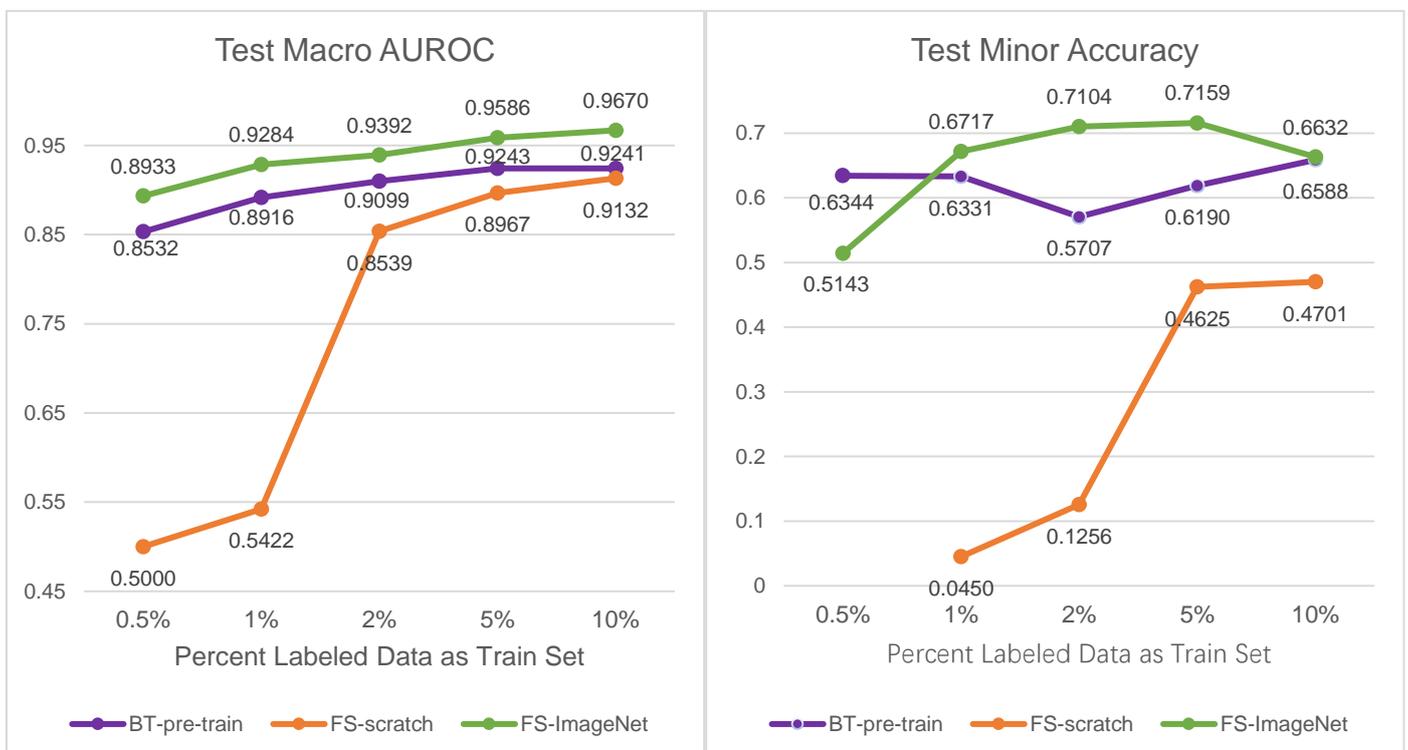

**Figure 2: Performance of BT and the supervised baseline**

In addition, we ablated types of distortion transformation during the BT pre-training process and discovered that removing any of them results in performance degradation. Even transformations such as solarization and rotations that produce less realistic transformations for histopathology images had a small but measurable impact on model performance (**Table 4, Figure 3**). All models presented in **Table 4** are pre-trained for 30 epochs instead of 100 to reduce computation time; results were obtained using the 10% training set. For BT-id, the model was given two batches of identical images. Since the BT loss function also encourages non-redundant features, the network trains despite having identical input images.

Interestingly, the BT-all model, which was pre-trained for 30 epochs, slightly outperforms BT-pre-train, which was pre-trained for 100 epochs. To further determine if training times can be reduced, we explored the performance of other BT models trained using 30 epochs. These experiments showed that BT-pre-train significantly outperforms BT-all using smaller datasets (e.g., at 0.5%, BT-pre-train achieves a macro AUROC of 0.8532, while BT-all only achieves 0.7037). However, at higher data percentages (10% and more), BT-all tends to match BT-pre-train in performance. We speculate that this is because the additional knowledge gains from a longer pre-training period ceases to be advantageous when significant amounts of labeled data are available for fine-tuning. For reference, the performance of the BT-all model fine-tuned on the 80% train set is provided in **Table 3**.

**Table 3: Performance of BT-all using 80% train set (30 epochs pre-training).**

| AUROC (micro) | 0.9718 |
|---|---|
| AUROC (macro) | 0.9495 |
| Minor Acc. | 0.7084 |

## 4.4 Robustness of BT representations and other observations

Conceptually, one of the appeals of Barlow Twins and similar methods is that by design they emphasize learning of robust representations through extensive data augmentation. Given the importance of pre-analytic variability in histology images and differences across medical labs, we hypothesized that this design feature might be beneficial for computational pathology, producing representations that are more domain-invariant than fully supervised methods.

We investigated this using t-SNE visualization of BT and ImageNet representations (**Figure 4**). We performed t-SNE on features from the BT pre-trained ResNet50 backbone and the ImageNet initialized backbone features. Instances were color coded by the hospital associated with each instance.

**Table 4: BT augmentations ablations**

| Method | AUROC (micro)* | AUROC (macro) | Minor Acc. |
|---|---|---|---|
| BT-all | 0.9646 | 0.9457 | 0.6664 |
| BT-no-drop | 0.9643 | 0.9212 | 0.6285 |
| BT-no-color-no-drop | 0.9354 | 0.9058 | 0.5211 |
| BT-blur-only | 0.9242 | 0.8796 | 0.5373 |
| BT-id | 0.9278 | 0.8743 | 0.4980 |

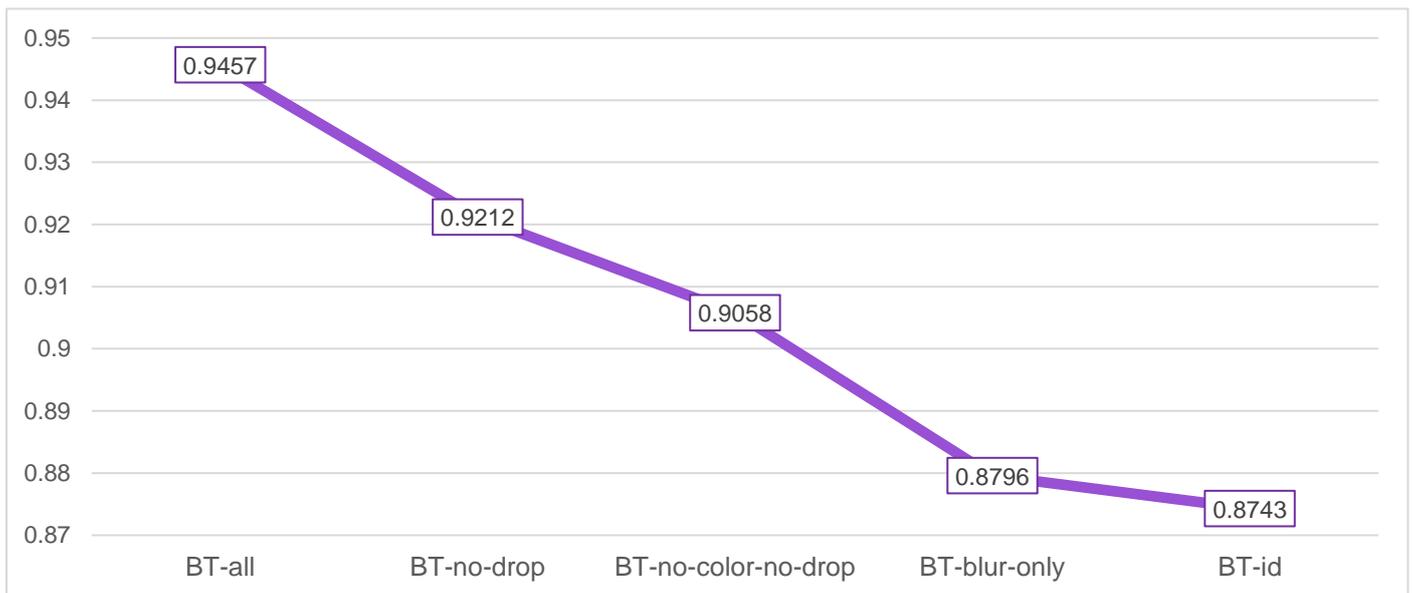

**Figure 3: BT augmentations ablations trend line**

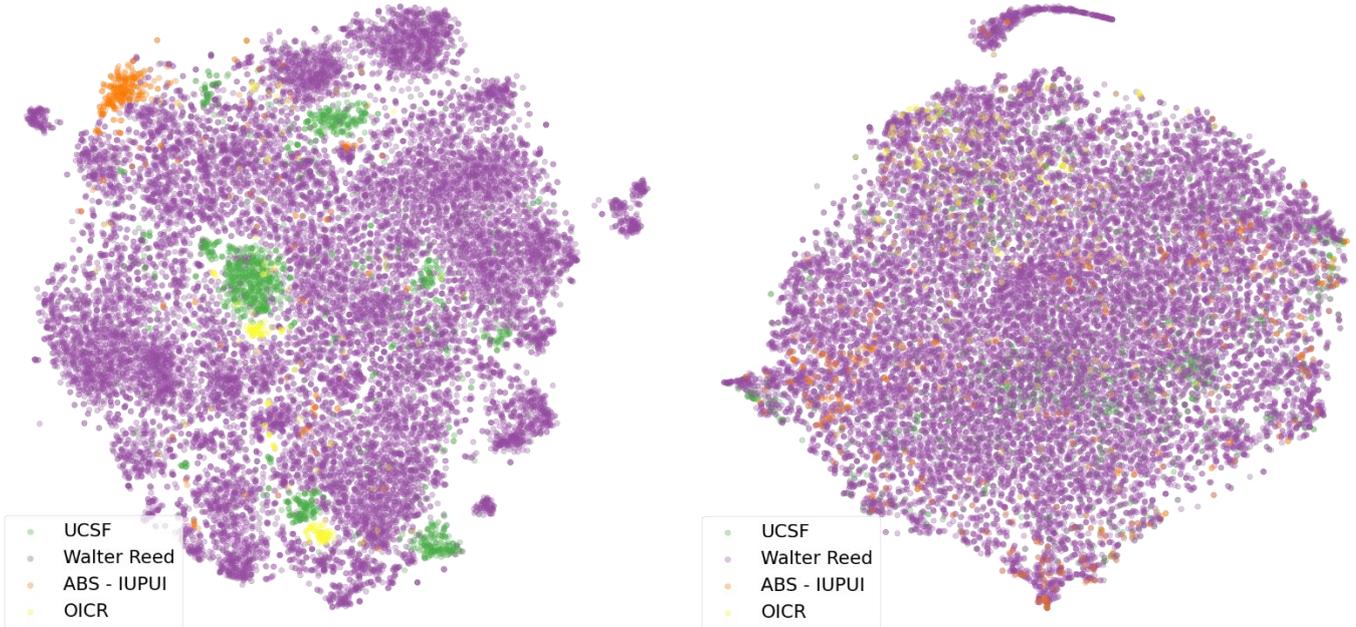

**Figure 4: Output feature clustering by testing set hospital using t-SNE.** Clustering is an adverse outcome signaling less generalization potential. BT initialization reduces clustering and is potentially more domain invariant.

In the ImageNet t-SNE plot there is significant clustering based on hospital, which is an undesired outcome signaling less out-of-domain generalization. In contrast, there is no clustering in the BT pre-trained plot. To quantify this observation, we trained a simple support vector machine (SVM) to classify BT and ImageNet representations based on test set hospitals. While the SVM achieves an accuracy of 97.94% for ImageNet representations, it only achieves 88.82% for BT representations, which is equivalent to random guessing since 88.96% of test data comes from Walter Reed hospital. Although the BT representation features are clearly more domain agnostic, the predictive accuracy in high-dimensional space of ImageNet provides a discriminative advantage.

We also observed differences in model behavior during training. The BT model seems to be naturally resilient to overfitting during the fine-tuning stage. Without using any additional regularization techniques, the performance gap between the validation and train datasets remains small throughout training. Even when the learning rate is intentionally increased, this behavior remains stable. In contrast, for both FS-scratch and FS-ImageNet, we observe significant overfitting of the training set at convergence via performance gaps on training and validation. We surmise that the extensive augmentation used by BT provides a certain level of regularization that prevents overfitting, which is consistent with our observation that features produced by the BT model are more domain-invariant. This robustness does not, however, translate to improved predictive accuracy in our tissue classification task.

## 5  CONCLUSIONS

We investigated BT using a simple breast cancer tissue classification dataset, and compared the performance of BT to alternative fully supervised models with and without ImageNet initialization. While BT can learn effective representations, the more conventional ImageNet representation had better performance in this task. Our observation that BT performs better than FS-scratch models supports a similar observation by Koohbanani *et al.* [8]. Their experiments did not, however, compare self-supervision to ImageNet initialized models like FS-ImageNet, and so more investigation is needed to compare the performance of SSL and FS models like FS-/image/net.

For training BT models in the context of histopathology, we observe that during the pre-training stage, the inclusion of more transformations is beneficial, even if they are potentially irrelevant to the final task. This is consistent with observations when applying irrelevant style transformations to histopathology for data augmentation [14]. We did not observe a significant effect on performance as we varied the dimensionality of the BT projector network. For e.g., the micro-AUROCs for BT models with 2048 and 4096 dimensionalities are 0.9720 and 0.9737 respectively. For smaller data percentages, models with larger dimensions tend to perform better, but the difference is minimal. Finally, the BT pre-training stage is especially effective at producing domain-invariant features and providing regularization during the fine-tuning process, as observed in our multi-institutional dataset.

Our study is not definitive and has important limitations. First, we examined only one task, and so the conclusions of this study do not necessarily generalize to all histopathology applications. Second, our baseline models did not utilize performance-improving techniques like data augmentation, and so the performance values cited are likely conservative. Future work should expand the applications examined, and investigate other SSL training paradigms such as auxiliary task learning [8].